\documentstyle[twocolumn,floats,aps,prd,overcite]{revtex}

\newcommand{\be}{\begin{equation}}
\newcommand{\ee}{\end{equation}}
\newcommand{\bea}{\begin{eqnarray}}
\newcommand{\eea}{\end{eqnarray}}
\newcommand{\bml}{\begin{mathletters}}
\newcommand{\eml}{\end{mathletters}}

\begin{document}
\preprint{DTP/99/87, hep-th/9912160}



\wideabs{                       
\title{ 
Wave function of the radion in a brane world
}
\author{Christos Charmousis$^1$, Ruth Gregory$^{1}$ and 
Valery A. Rubakov$^{2,3}$}

\address{~$^1$ Centre for Particle Theory, Durham University,
South Road, Durham, DH1 3LE, U.K.\\
{~}$^2$ Institute for Nuclear Research of the Russian Academy of Sciences,\\
60th October Anniversary prospect, 7a, Moscow 117312, Russia.\\
{~}$^3$ Isaac Newton Institute for Mathematical Sciences,\\
University of Cambridge, 20 Clarkson Road, Cambridge, CB3 0EH, U.K. 
}

\maketitle
\begin{abstract}
We calculate the linearized metric perturbation corresponding
to a massless four-dimensional scalar field, the radion, in a
five-dimensional two-brane model of Randall and Sundrum.
In this way we obtain relative strengths of the radion couplings to 
matter residing on each of the branes. The results are in agreement
with the analysis of Garriga and Tanaka of gravitational and
Brans--Dicke forces between matter on the branes. We also introduce a 
model with infinite fifth dimension and ``almost'' confined graviton,
and calculate the radion properties in that model.

PACS numbers: 04.50.+h, 11.25.Mj \hfill hep-th/9912160
\end{abstract}
}                               


Recently, considerable interest has been raised by a
five-dimensional model with an $S^1/Z_2$ orbifold extra dimension
with two 3-branes residing at its boundaries \cite{Randall:1999ee}.
This model and its non-compact analogues 
\cite{Gogberashvili:1999ad,Randall:1999vf,Cohen:1999ia,Gregory:1999gv}
(see Ref.\ \cite{Visser:1985qm} for an account of earlier works)
provide a novel setting for discussing various conceptual and
phenomenological issues related to compactification of extra dimensions 
in models motivated by M-theory. In the two-brane
Randall--Sundrum model
\cite{Randall:1999ee}, the  branes have tensions $+\sigma$ and
$-\sigma$, and the bulk cosmological constant is chosen in such a way 
that the classical solution describes five-dimensional space-time whose
four-dimensional slices are flat,
\be
ds^2 = a^2(z) \eta_{\mu\nu} dx^{\mu} dx^{\nu} - dz^2
\label{2*}
\ee
Here $a(z)=\mbox{e}^{-k|z|}$,
the fifth coordinate $z$ runs from $z_{+} = 0$ to $z_{-} = r_c$
and $k=(4\pi/3)G_5 \sigma$ where $G_5$ is Newton's
constant in five dimensions. The orbifold symmetry, a local reflection 
symmetry at each brane, is assumed to hold for all fields in this space-time.

The excitations above the background metric (\ref{2*}) contain
a massless four-dimensional
graviton (whose wave function is peaked at the positive tension
brane) and the corresponding Kaluza--Klein tower \cite{Randall:1999vf}. 
This is not the whole story, however. In general when one has a wall 
in spacetime, one might expect a translational zero mode giving
rise to free motion of the wall. In the case of anti-de Sitter 
spacetime, $\partial_z$ is not a translational Killing 
vector but a conformal Killing vector, nonetheless we can identify
solutions to the perturbation equations which correspond to proper motion
of the wall (although these will be singular on the AdS horizon). In
the conventional application of the Israel equations, one identifies 
the extrinsic curvature, $K_{\mu\nu}$,
on each side of the wall, and then applies a $Z_2$
symmetry across the wall leading to ${\tilde K}_{\mu\nu} =
(K^+_{\mu\nu} + K^-_{\mu\nu})/2 = 0$; geometrically this means that the
wall is locally `flat' i.e.\ totally geodesic. To describe proper dynamical
motion of the wall, we require a nonzero ${\tilde K}_{\mu\nu}$, which is
possible if the $Z_2$-symmetry is not imposed.
The appropriate solution for such a motion then turns out to be
\be
\delta g_{\mu\nu} = {\sinh 2kz\over2k} {\tilde K}_{\mu\nu} 
\label{propmot} 
\ee
where ${\tilde K}_\mu^\mu = 0$, which 
is recognised as the `Nambu' equation for a brane.
Since this solution blows up at large $z$, it does not correspond
to a small perturbation of the spacetime, and is indicative that in
the presence of such free motion, the asymptotic structure of the spacetime
is altered, similar to the difference between the metrics of a straight
cosmic string and a crinkly cosmic string \cite{Vil}. As such, this
perturbation is not considered in the general spectrum of localised
perturbations of the Randall--Sundrum wall.

Once we have two branes
however, the situation is different: there are now two sorts of motion, 
a centre of mass (which will still be divergent) and relative motion
-- the radion, for which the second wall acts as a regulator
on the divergence of (\ref{propmot}). It is this second mode which we wish to
identify,  which will correspond to a massless 
four-dimensional scalar. One may or may not suspect that the radion is 
also accompanied by its own Kaluza--Klein tower.

Naively, the radion field $T(x)$ has been introduced by
considering metrics of the form 
\cite{Randall:1999ee,Randall:1999vf,Csaki:1999mp,Goldberger:1999un}
\be
ds^2 = \mbox{e}^{-2k|z|T(x)} g_{\mu\nu}(x) dx^{\mu} dx^{\nu}
- T^2(x) dz^2   
\label{3*}
\ee
where $g_{\mu\nu}$ is the four-dimensional graviton. However, the
Ansatz (\ref{3*}) does not solve even the linearized field equations
that are obtained by setting $T(x) = 1 + f(x)$,  
$g_{\mu\nu} = \eta_{\mu\nu} + \chi_{\mu\nu}(x)$ with small $f$ and
$\chi_{\mu\nu}$. Furthermore, the Ansatz (\ref{3*}) would imply that the
radion does not interact with matter living on the positive tension brane.
The latter feature, unnatural by itself, would be in contradiction with 
the results of Garriga and Tanaka \cite{Garriga:1999yh},
who calculated the long-range forces between matter placed on the
branes and showed, in particular, that matter on each of the branes 
participates in interactions of the Brans--Dicke type.

The purpose of this note is to clarify this issue by
calculating, in linearized theory,
the five-dimensional metric perturbation corresponding to the
propagating radion field.
We will see that this perturbation does not vanish on either
of the branes. We will point out
also that there is no Kaluza--Klein tower above the 
radion, i.e., that all massive states have been accounted for
in the analysis of Ref.\cite{Randall:1999vf}.

To deal with the $Z_2$-symmetry as well as with junction conditions on 
the branes, it is convenient to choose Gaussian Normal (GN) coordinates
\be
g_{zz} = -1\,,\,\,\, g_{z\mu} = 0   
\label{5*}
\ee
Such a system can always be chosen in the neighborhood 
of the brane by integrating out along its normal, in which case
$z$ will be the proper distance from the brane. However, note that
this system is slightly more general, in that we can make coordinate
transformations which shift the wall, but preserve the metric 
components (\ref{5*}).

Then the linearized theory is described by the metric
\be
ds^2 = a^2(z) \eta_{\mu\nu} dx^{\mu} dx^{\nu} 
+ h_{\mu\nu}(x,z)dx^{\mu}dx^{\nu} - dz^2  
\ee
We will explicitly consider the region $r_c > z>0$; the orbifold
symmetry giving $h_{\mu\nu}$ for other values of $z$.
The four-dimensional indices will be raised and lowered using the
Minkowski metric $\eta_{\mu\nu}$.
The linearized Einstein equations are
\bml\bea
\delta R_{zz} &=& 8\pi G_5\left( \frac{2}{3} T_{zz} + 
\frac{1}{3a^2} T^{\lambda}_{\lambda} \right) \\
\delta R_{z\mu} &=& 8\pi G_5 T_{z\mu} \\
\delta R_{\mu\nu} - 4k^2 h_{\mu\nu} &=& 8\pi G_5\left( T_{\mu\nu}
- \frac{1}{3} \eta_{\mu\nu}T^{\lambda}_{\lambda} + 
\frac{a^2}{3} \eta_{\mu\nu} T_{zz} \right)
\eea\eml
Here $T_{ab}$ is the energy-momentum tensor of additional matter,
if present, and
\bml\bea
\delta R_{zz} &=& - \left(\frac{h'}{2a^2}\right)' 
-2k\delta(z) h + {2k\over a^2(r_c)} \delta(z-r_c) h \\
\delta R_{z\mu} &=& \left(\frac{1}{2a^2}(h^{\nu}_{\mu,\nu}
-h_{,\mu} )\right)' \nonumber \\
\delta R_{\mu\nu}&=&\frac{1}{2}h_{\mu\nu}'' + 2k^2 h_{\mu\nu}
-\left(k^2 h + \frac{k}{2} h'\right)\eta_{\mu\nu} \nonumber\\ 
&& + 2k \left (\delta(z) - \delta(z-r_c) \right ) h_{\mu\nu} \nonumber \\
&&  +\frac{1}{2a^2} (2 h^{\lambda}_{(\mu,\nu)\lambda} 
- h^{\;\;\;\;\lambda}_{\mu\nu,\lambda} - h_{,\mu\nu} )
\eea \label{6*} \eml
where $h = h^{\mu}_{\mu}$.
Equations (\ref{6*}) are invariant under residual gauge transformations,
\be
h_{\mu\nu} \to  h_{\mu\nu} + a^2(\epsilon_{\mu,\nu} + \epsilon_{\nu,\mu})
+ \frac{1}{k} \epsilon^z_{\;,\mu\nu} - 2ka^2 \eta_{\mu\nu}\epsilon^z   
\label{6++}
\ee
where $\epsilon^z$ and $\epsilon^{\mu}$ depend only on $x$.
These transformations correspond to general coordinate
transformations $z \to z + \xi^z$, $x^{\mu} \to x^{\mu} + \xi^{\mu}$;
their consistency with the gauge conditions (\ref{5*}) requires
$\xi^z = \epsilon^z (x)$, 
$\xi^{\mu} = (2k)^{-1}a^{-2} \epsilon^{z,\mu}(x) + \epsilon^{\mu}(x)$.

The Israel junction conditions on a brane are most easily 
formulated in the GN frame, in which the brane is located
at fixed $z$. In the absence of matter on the brane, these junction 
conditions are $h_{\mu\nu}' + 2k h_{\mu\nu} = 0$. 
They
are {\it not} invariant under the gauge
transformations (\ref{6++}) if $\epsilon^z \neq 0$. The importance 
of the gauge transformations (\ref{6++}) becomes clear from the 
observation that the coordinate system which is GN
with respect to one brane need not be GN with respect
to the other. 
Hence, one is led to consider two coordinate patches,
the first (second) of which includes the positive (negative) tension
brane. The coordinate systems in each of these patches are GN
to the respective brane. A residual coordinate transformation is
needed then to relate the metrics in the overlap of these patches.

In other words, to describe the propagating degrees of freedom, 
we introduce two sets of fields, $h^{(+)}_{\mu\nu}(x,z)$ and
$h^{(-)}_{\mu\nu}(x,z)$. The first of them,  $h^{(+)}_{\mu\nu}$,
is defined in the interval in the fifth direction that includes 
$z_{+}=0$ but excludes  $z_{-} = r_c$,
and conversely for  $h^{(-)}_{\mu\nu}$. Both  $h^{(\pm)}_{\mu\nu}$ obey
source-free equations (\ref{6*}). The boundary conditions are 
\be
h^{(\pm)\prime}_{\mu\nu} +2kh^{(\pm)}_{\mu\nu} = 0 
\,\,\,\mbox{at} \,\,\, z=z_{\pm}
\label{jumppm}
\ee
The relation between the two fields in the bulk is the gauge 
transformation of the form (\ref{6++})
with yet unknown gauge functions.

For $T_{ab}=0$, the linearized Einstein equations (\ref{6*}) 
with boundary conditions (\ref{jumppm}) are 
straightforward to solve by ``brute force''.
The outcome can be understood as follows.
In each of the patches we write
\be
h^{(\pm)}_{\mu\nu} = \tilde{h}^{(\pm)}_{\mu\nu} +
\frac{1}{k} f^{(\pm)}_{,\mu\nu} - 2ka^2 \eta_{\mu\nu}f^{(\pm)}
\ee
where $f^{(\pm)}(x)$ are yet to be determined and 
$\tilde{h}^{(\pm)}_{\mu\nu}$ is transverse-tracefree (TT)
$\tilde{h}^{(\pm)\mu}_{\mu} = 0$,
$\tilde{h}^{(\pm)\nu}_{\mu,\nu} =0$.
Then the field equations in the bulk become
\be
\tilde{h}^{(\pm)\prime\prime}_{\mu\nu}
-2k^2 \tilde{h}^{(\pm)}_{\mu\nu}
-\frac{1}{2a^2} \Box^{(4)}
\tilde{h}^{(\pm)}_{\mu\nu} = 0
\label{9*}
\ee
while the junction conditions on the respective branes read
\be
\tilde{h}^{(\pm)\prime}_{\mu\nu} +2k \tilde{h}^{(\pm)}_{\mu\nu}
= -2 f^{(\pm)}_{,\mu\nu}    
\label{9**}
\ee
The latter are consistent with TT-property of $\tilde{h}_{\mu\nu}$
iff $\Box^{(4)} f^{(\pm)} = 0$.
Hence, if the four-dimensional momenta are such that $p^2\neq 0$,
one is left with eq.(\ref{9*}) and homogeneous boundary conditions
(i.e., eq.(\ref{9**}) with $f^{(\pm)} = 0$). This is precisely the 
system of equations analyzed in Ref.\cite{Randall:1999vf}, so we
see that all massive propagating modes have been
revealed by that analysis.

At $p^2=0$, however, there are two types of solutions. One of them 
is 
$f^{(\pm)}(x)= 0$, $\tilde{h}_{\mu\nu}(x,z) = a^2\chi_{\mu\nu}(x)$, 
and does not 
require the gauge transformation in the overlap of the two patches.
These solutions have  been considered
in Ref.\cite{Randall:1999vf}, and describe
massless four-dimensional gravitons. The other type of solution
is
\be
\tilde{h}^{(\pm)}_{\mu\nu} = - \frac{a_{\pm}^2}{2ka^2}  
f^{(\pm)}_{,\mu\nu}      
\ee
where $f^{(\pm)}(x)$ are yet arbitrary 
and
$a_{\pm} = a (z_{\pm})$. The relation between $f^{(+)}$ and $f^{(-)}$
is found using (\ref{6++}). Since $ \tilde{h}_{\mu\nu}$
are proportional to $a^{-2}$, they should coincide in the two patches,
so one requires $a_{+}f^{(+)} (x) =  a_{-}f^{(-)} (x) \equiv f(x)$.
One obtains finally
\bml\bea
h^{(+)}_{\mu\nu} &=& -\frac{1}{2ka^2} f_{,\mu\nu}
+ \frac{1}{k} f_{,\mu\nu} -2ka^2\eta_{\mu\nu} f \label{11*}\\
h^{(-)}_{\mu\nu} &=& -\frac{1}{2ka^2} f_{,\mu\nu}
+ \frac{\mbox{e}^{2kr_c}}{k} f_{,\mu\nu}
-2ka^2\mbox{e}^{2kr_c}\eta_{\mu\nu} f
\label{11**}
\eea\eml
where $f(x)$ is a massless four-dimensional scalar mode.
The first term on the RHS is clearly identifiable as the growing part of
the mode in eq.\ (\ref{propmot}), and hence corresponds to motion
of the wall; the coincidence of this term in (\ref{11*}) and
(\ref{11**}) identifies this as a relative motion. One can
quantify this by noting that the
transition function between the two patches is
\be
\epsilon^z (x) = \left( \mbox{e}^{2kr_c} -1 \right) f(x) \,,
   \,\,\, \epsilon_{\mu} = 0
\label{add2}
\ee
This transition function then determines the physical distance between the
branes, $r(x) - r_c = \epsilon^z (x)$
(recall that eq.\ (\ref{6++}) is the coordinate transformation
between the coordinate systems in which the branes are located
exactly at $z=0$ and $z=r_c$, respectively). These properties
show that $f(x)$ is indeed the (unnormalized)
radion field in the linearized theory.

Equations (\ref{11*}) and (\ref{11**}) determine the induced metrics
on each of the branes in the presence of the radion field.
The first two terms on the right hand sides of these equations can be
gauged away {\it on the branes}. With the graviton field 
$\chi_{\mu\nu}$  included, the induced metrics on each brane are
\be
\bar{h}^{(\pm)}_{\mu\nu}(x)
= a^2_{\pm} \left(\eta_{\mu\nu} + \chi_{\mu\nu}(x)
- \frac{2k}{a^2_{\pm}} f(x) \eta_{\mu\nu} \right)    
\label{12*}
\ee

If matter is present on the branes, it couples to the induced metrics
through $L_{int}\propto \bar{h}_{\mu\nu}T^{\mu\nu}$. 
Clearly, the radion field couples to
the trace of energy-momentum tensor. The corresponding effective coupling 
constants at each brane are proportional to
$g_{\pm} \propto a_{\pm}^{-1}$.
Indeed, the elementary vertex of a graviton to matter at each brane
is proportional to $\sqrt{G_{N\pm}}$ where $G_{N\pm} \propto a^2_{\pm}$ 
are  effective four-dimensional Newton's constants at each 
brane \cite{Randall:1999ee}; from eq.(\ref{12*}) it follows that
the radion vertex contains an extra  factor $a^{-2}_{\pm}$.
Hence, the radion field couples to matter on the
negative tension brane exponentially stronger than to matter on
the positive tension one, $g^2_{-}/g^2_{+} = \mbox{e}^{2kr_c}$.
This relation is just the opposite to the case of graviton, and
it is in accord with the results of Garriga and Tanaka 
\cite{Garriga:1999yh}. The overall 
strength of these interactions can be also read off from 
Ref.\cite{Garriga:1999yh}:  the interaction Lagrangian of the
normalized 
radion field $\hat{f}(x)$ with matter on each brane is
$L_{int}^{(\pm)} = g_{\pm} \hat{f} T^{(\pm)\mu}_{\mu}$
with
\be
g^2_{\pm} = \frac{16\pi}{3}G_5 k
\frac{\mbox{e}^{\mp kr_c}}{\sinh kr_c}   
\ee
The fact that the radion couples to matter on the negative tension 
brane much stronger than the graviton does has been observed also in
Refs.\cite{Csaki:1999mp,Goldberger:1999un}.

It is instructive to return to (\ref{3*}) with the benefit
of our perturbative calculation to see what the linearised metric
with the walls fixed  at some coordinate values, $0$ and $r_c$, should
look like. To derive this form, we take the two GN patches
of (\ref{11*}) and (\ref{11**}), and perform a gauge transformation
to make the two identical. 
We now have a single 
coordinate chart between the walls, but the walls are no longer
at $z = 0, r_c$. We then perform another coordinate transformation
which is determined by the dual requirements that the walls sit at the
(new) $\tilde z$ coordinates $0$ and $r_c$, and that there are no
cross terms ${\tilde g}_{{\tilde z}\mu}$ in the metric. The price of
having the walls at a rigid value of $\tilde z$ is that the system
is no longer GN -- a nontrivial ${\tilde g}_{{\tilde z}{\tilde z}}$ is 
introduced. 
After performing these transformations we find that the new metric
is
\be
d{\tilde s}^2 = \mbox{e}^{-2k({\tilde z} + f(\tilde{x})e^{2k{\tilde z}})}
g_{\mu\nu}({\tilde x}) d{\tilde x}^\mu d{\tilde x}^\nu - \left ( 1 + 
2kf(\tilde{x})
\mbox{e}^{2k{\tilde z}} \right )^2 d{\tilde z}^2
\label{newnew}
\ee
where we have included the possibility of graviton perturbations
in $g_{\mu\nu}({\tilde x})$. 
Unlike eq. (\ref{3*}), this form of the metric correctly describes both 
the linearized dynamics of massless fields and 
$\tilde{x}$-independent displacements of the wall (the latter correspond 
to constant, but not necessarily small $f$ and are accompanied by
a change of $\tilde{z}$ coordinate).
Therefore, eq.(\ref{newnew}) is expected to appropriately describe the
full long distance dynamics of  the two brane model.

Clearly, the properties of the radion  are quite different
from  the graviton. To stress this point, let us introduce a model
in which gravitons are not confined, but the radion is. This model may 
be of interest by itself, as in an appropriate limit gravity on a
brane is expected to be almost, but not exactly Einsteinian.

Let us consider five-dimensional space-time with infinite
fifth dimension. Let there be two branes, one with positive tension
$\sigma$ and another with negative tension $-\sigma/2$
(note the factor $1/2$). The latter brane
is placed to the right of the former in the fifth direction.
The bulk cosmological constant between the two branes and to the 
left of the positive-tension one is the same as in the
Randall--Sundrum model, and is zero to the right of the 
negative-tension brane. Then there exists a solution to the
Einstein equations for which both branes are at rest, the coordinates
of the positive and negative tension branes being $z=0$ and $z=r_c$,
respectively, where $r_c$ is again an arbitrary constant. 
This solution has the form of eq.(\ref{2*}) but now with
\be
a^2(z) = \cases{ \mbox{e}^{-2k|z|}& for $z<r_c$\cr
\mbox{e}^{-2kr_c} = \mbox{const} & for $z>r_c$\cr}
\label{2a+}
\ee
The four-dimensional hypersurfaces $z=\mbox{const}$ are flat;
the five-dimensional space-time is flat to the right of the
negative tension brane, and anti-de Sitter in the rest of the bulk.

An interesting feature of this model is that gravitons are almost
but not exactly confined:  the wave functions of gravitons,
$h_{\mu\nu} = a^2(z)\chi_{\mu\nu}(x)$,
are peaked at $z=0$ but are not normalizable. 
At large $r_c$, gravity experienced by matter residing
on the positive tension brane should be almost, but not exactly
Einsteinian (the limit $r_c \to \infty$ corresponds to the
non-compact GRS model \cite{Gogberashvili:1999ad,Randall:1999vf},
with gravitons confined to the positive tension brane). The background
(\ref{2a+}) is of interest for  exploring possible deviations from the
Einstein gravity in brane world, and, in particular, for analyzing
the issue of (non) conservation of energy measured by a four-dimensional 
observer.

We leave the discussion of gravitational perturbations in our model 
for the future, and here we consider a simpler mode, the radion.
For the confined radion, the metric perturbation analogous to
eqs.(\ref{11*}), (\ref{11**}) has to be a solution to linearized
Einstein equations (still in the gauge (\ref{5*})) which tends to pure
gauge as $z \to +\infty$ and $z \to -\infty$. We again have to consider 
two coordinate patches,
overlapping in a region between the branes.
In the overlap, the perturbations $h^{(+)}_{\mu\nu}$ and
$h^{(-)}_{\mu\nu}$ are to be related by a gauge transformation
(\ref{6++}). 

Proceeding as above, we find in the left patch
\be
h^{(+)}_{\mu\nu} = \frac{1}{k}\left( 1 - \mbox{e}^{2kz}\right) f_{,\mu\nu} (x)
- 2 k a^2 f(x) \eta_{\mu\nu}
\label{5a*}
\ee
where
$f(x)$ is the massless radion field. 
The forms of metric perturbation
in the right patch are different in anti-de Sitter and flat 
parts,
\be
h^{(-)}_{\mu\nu} = \cases{
\frac{2}{k} \mbox{e}^{2kr_c} \sinh[2k(r_c-z)] f_{,\mu\nu}
& for $z<r_c$, \cr
- 4 (z-r_c)  \mbox{e}^{2kr_c}   f_{,\mu\nu} & for $z>r_c$. \cr}
\label{rtpatch}
\ee
It is straightforward to see that these perturbations indeed
obey the Einstein equations everywhere in the bulk and the
Israel junction conditions on the branes. The gauge
transformation relating
$h^{(+)}_{\mu\nu}$ and
$h^{(-)}_{\mu\nu}$
in the bulk between the two branes is
\begin{equation}
h^{(+)}_{\mu\nu} - h^{(-)}_{\mu\nu} = \frac{1}{k} f_{,\mu\nu} 
-2ka^2 f \eta_{\mu\nu} -\frac{1}{k} \mbox{e}^{4kr_c} a^2 f_{,\mu\nu} 
\end{equation}
i.e., in the notation of eq.\ (\ref{6++})
\be
\epsilon^z = f \;\;\; ; \;\;\; \epsilon_\mu = {f_{,\mu}\over2k} 
\left ( 1 - \mbox{e}^{4kr_c} a^2 \right )
\ee
It is easy to see that this corresponds to proper relative
motion of the wall, since computing the extrinsic curvature of
the first wall gives ${\tilde K}^{(+)}_{\mu\nu} = - f_{,\mu\nu}$.
Meanwhile, at the second wall 
${\tilde K}^{(-)}_{\mu\nu} = -2e^{2kr_c} f_{,{\mu}{\nu}}$.  Alternatively,
the perturbation (\ref{rtpatch}) is pure gauge for $z>r_c$, and
changing coordinates to the right of the second wall so that
the metric there is Minkowskian, we find that the wall is located
at ${\hat z}^{(-)}= r_c-2\mbox{e}^{2kr_c}f$.  Similarly, for
$z<0$, the perturbation (\ref{5a*}) is pure gauge, and changing
coordinates for the first wall gives $\hat{z}^{(+)} = -f$, 
therefore we see
how $f$ does indeed encode a relative motion of the walls.
Note how the radion field is non-trivial only in between the two branes
and on the positive tension brane itself. In other words, there is not even
short-ranged radion hair outside the two-brane system. 
It is likely that the absence of the radion hair
outside a stack of branes is 
a general property of models with infinite extra 
dimensions.

Finally, we note that in our model
the radion does not induce metric perturbations
on the negative tension brane, $ h^{(-)}_{\mu\nu} (r_c) = 0$.
Hence, the radion does not interact with matter residing on the
negative tension brane, in sharp contrast to the Randall--Sundrum model
discussed above. This seems to be a peculiarity of our model, which is
related to the flatness of the five-dimensional
space-time for $z>r_c$, since a perturbed wall in flat spacetime written
in GN coordinates can be shown to have four-dimensional metric
$g_{\mu\nu} = \eta_{\mu\nu} + 2z f_{,\mu\nu} + O(f^2)$,
and so any perturbation always vanishes to leading order on the wall.

\section*{Acknowledgements}

We would like to thank Jaume Garriga, Takahiro Tanaka, David Langlois
and Clifford Johnson for useful discussions. This work was done while
participating in the workshop on {\it `Structure Formation in the
Universe'} at the Isaac Newton Institute. C.C.\ was supported by PPARC,
R.G.\ by the Royal Society, and V.R. by
the Russian Foundation for Basic Research, grant 990218410.

\def\apj#1 #2 #3.{{\it Astrophys.\ J.\ \bf#1} #2 (#3).}
\def\cmp#1 #2 #3.{{\it Commun.\ Math.\ Phys.\ \bf#1} #2 (#3).}
\def\comnpp#1 #2 #3.{{\it Comm.\ Nucl.\ Part.\ Phys.\  \bf#1} #2 (#3).}
\def\cqg#1 #2 #3.{{\it Class.\ Quant.\ Grav.\ \bf#1} #2 (#3).}
\def\jmp#1 #2 #3.{{\it J.\ Math.\ Phys.\ \bf#1} #2 (#3).}
\def\ijmpd#1 #2 #3.{{\it Int.\ J.\ Mod.\ Phys.\ \bf D#1} #2 (#3).}
\def\mpla#1 #2 #3.{{\it Mod.\ Phys.\ Lett.\ \rm A\bf#1} #2 (#3).}
\def\ncim#1 #2 #3.{{\it Nuovo Cim.\ \bf#1\/} #2 (#3).}
\def\npb#1 #2 #3.{{\it Nucl.\ Phys.\ \rm B\bf#1} #2 (#3).}
\def\phrep#1 #2 #3.{{\it Phys.\ Rep.\ \bf#1\/} #2 (#3).}
\def\pla#1 #2 #3.{{\it Phys.\ Lett.\ \bf#1\/}A #2 (#3).}
\def\plb#1 #2 #3.{{\it Phys.\ Lett.\ \bf#1\/}B #2 (#3).}
\def\pr#1 #2 #3.{{\it Phys.\ Rev.\ \bf#1} #2 (#3).}
\def\prd#1 #2 #3.{{\it Phys.\ Rev.\ \rm D\bf#1} #2 (#3).}
\def\prl#1 #2 #3.{{\it Phys.\ Rev.\ Lett.\ \bf#1} #2 (#3).}
\def\prs#1 #2 #3.{{\it Proc.\ Roy.\ Soc.\ Lond.\ A.\ \bf#1} #2 (#3).}

\end{document}